\documentclass[11pt]{article}
\usepackage{graphicx, caption, subfigure}
\makeatletter
\newcommand{\singlespacing}{\let\CS=\@currsize\renewcommand{\baselinestretch}{1.0}\tiny\CS}
\newcommand{\doublespacing}{\let\CS=\@currsize\renewcommand{\baselinestretch}{1.5}\tiny\CS}

\begin{document}
\title{Direct Photon Production at RHIC and LHC-energies: Measured Data Versus a Model}
\author{P.
Guptaroy$^1$\thanks{e-mail: gpradeepta@rediffmail.com (Communicating author)}, S. Guptaroy$^2$\thanks{e-mail: simaguptaroy@yahoo.com}\\
{\small $^1$ Department of Physics, Raghunathpur College,}\\
 {\small P.O.: Raghunathpur 723133,  Dist.: Purulia (WB), India.}\\
 {\small $^2$ Department of Physics, Basantidevi College,}\\
 {\small 147B Rashbehari Avenue, Kolkata 700029 India.}}
 \date{}
\maketitle
\begin{abstract}
We attempt, in this paper, to deal with the direct photon production phenomena in $p+p$, $d+Au$ and $Au+Au$ collisions at RHIC energy, $\sqrt{s_{NN}}$ = 200 GeV and in $Pb+Pb$-collisions at LHC energy, $\sqrt{s_{NN}}$ = 2.76 TeV  on the basis of a non-standard framework outlined in
the text. Comparisons of the
model-based results with the measured data on some observables reveal fair agreement; and they are modestly consistent with the results obtained by some other models of `Standard' framework.
\end{abstract}
\newpage
 One of the major goals in studying the collisions
between two nuclei at RHIC and of LHC energies is the observation of formation of quark gluon
plasma (QGP). But, till date, the exact nature of QGP-
hadron phase transition is still plagued by uncertainties.\cite {sinha} Direct photon production have long been recognised as one of the important signals of the QGP-phase. The reasons are as follows: (i) By
both ascription and definition, direct photons are those photons which are not produced by the decays of either neutral pions or eta-mesons; (ii) Secondly, once produced they hardly ever interact; and so they leave the system with their energy and momenta unaltered.
\par
According to the Standard Model (SM), direct photons are originated either by Compton process ($qg \rightarrow q\gamma$) or by annihilation process ($q \bar{q} \rightarrow g\gamma$). Direct single photon data are therefore very
sensitive to gluons in the SM-based approaches. But, in our approach, instead of using these two mechanisms for the production of the direct photon, we try to use an alternative model for the production of direct photon which is generally a non-standard type and is known as Sequential Chain Model (SCM). Our objectives are: (i) to interpret the latest data of direct photon production at RHIC and LHC with the help of this SCM and compare it with some other standard approaches; and (ii) to put an alternative approach for understanding the nucleonic structure and the reaction mechanism at high energy nuclear collisions.
\par
The salient features and the important physical characteristics of the SCM valid for hadron-hadron collisions are as follows: \cite{pgr031}-\cite{pgr08} (i) according to
this model, high energy
hadronic interactions boil down, essentially, to the pion-pion
interactions; as the protons are conceived in this model as
$p$=($\pi^+$$\pi^0$$\vartheta$), where $\vartheta$ is a spectator
particle needed for the dynamical generation of quantum numbers of
the nucleons. (ii) the incident energetic $\pi$-meson emits a $\rho$-meson, the $\rho$-meson then emits a $\pi$-meson and is changed into a $\omega$-meson, the $\omega$-meson emits a photon, which is called the direct photon ($\gamma_d$) form the viewpoint of the present model, and the $\pi$-meson is transformed once again into $\rho$-meson and the process of direct photon production continues. The schematic diagram of the direct photon production is shown in the Figure 1.
\par
The model, applied here, does not admit any compartmentalisation of ``soft" (low $p_T$) and ``hard" (large $p_T$, $p_T \geq 2$ GeV/c) production. Rather it presents a unified approach. Besides, the fundamental expressions for final calculations are derived here on the basis of field-theoretic considerations and Feynman diagram technique with the infinite momentum frame tools and under impulse approximation method. On the whole, the SCM is purely analytical approach with a reasonable number of valid assumptions and approximations. Moreover, the various coupling strengths used in this model are well known and well measured by several experiments. This is one of very strong points about this model. The model can accommodate a large amount of cosmic ray physics issues and it can explain a majority of the characteristics of what are known to be the ``quark-gluon plasma"-hypothesis.
\par
The main form of the working formula is physically controlled by three factors
here: (i) production of the $\rho-\omega-\pi$
mesons in a Sequential chain and in cascade pattern in $pp$ collisions and the subsequent
production of direct photons by the free, emitted $\rho^0$ and
$\omega^0$ mesons, thus giving rise to the resemblance with a sort of
vector-meson-dominance approach; (ii) introduction of structural rearrangement
factors for large $p_T$ reactions studied here; (iii) induction of the nucleus
dependence by the Glauberian techniques.
\par
With the roles to be played by this triad, we obtain the nature of
the $p_T$ spectra for direct photon production at different interactions at heavy ion collisions.
\par
The basic
expression for inclusive cross-section for production of direct
photon in proton-proton collisions at high transverse momentum in the light of the SCM is \cite{sb88}
\begin{equation}\displaystyle{
E{\frac{d^3\sigma}{dp^3}}|_{pp \rightarrow \gamma_d X} \simeq
C_{\gamma_d}{\frac{1}{(p_T)^{N_R}}} {(\frac{1000}{s})}^{1/5}
\exp{(-65<n_{\gamma_d}>_{pp} x_T)},}
\end{equation}
where, for example, $C_{\gamma_d} \simeq 0.7$ for energy region of
Intersecting Storage Ring (ISR) experiments and it is different
for different energy regions. Besides, the terms $s$, $p_T$
represent respectively the square of the c.m. energy and the
transverse momentum, with $x_T = 2p_T/\sqrt s$, and
$<n_{\gamma_d}>$ the average multiplicity of direct photon. The
average multiplicity of the direct photon is deduced to be given
by the undernoted relation
\begin{equation}\displaystyle{
<n_{\gamma_d}>_{pp} ~ = ~ [\frac{{f^2_{\rho \omega \pi}}{g^2_{\omega {\gamma_d}}} (p^2_{\gamma_d})_{max}}{64 \pi^2}]^{1/3} ~ \simeq ~ <n_{\pi^0}>_{pp}(\frac{\alpha}{15})^{1/3},}
\end{equation}
In Eq. (2) given above, $\alpha$ is the fine structure constant. $f^2_{\rho \omega \pi}/4 \pi$ and $ g^2_{\omega {\gamma_d}}/4\pi$ are the coupling strengths for $\rho \omega \pi$ and for $\omega {\gamma_d}$ coupling respectively. And the factor $(p^2_{\gamma_d})_{max}$ is the sum of the square of the momentum for emitted individual direct photon which is finally expressed in terms of c.m. energy of the system with some simplifying high-energy approximations and assumptions. $<n_{\pi^0}>_{pp}$ in eqn.(2) represents the average multiplicity of neutral pions  and is $\simeq ~ 1.1s^{1/5}$.
\par
The second term in the right hand side of the equation (1),  $1/p_T^{N_R}$, the
constituent rearrangement term arises out of the partonic
rearrangements inside the proton. These
rearrangements mean undesirable loss of energy , in so far as the
production mechanism is concerned. The choice of ${N_R}$ would
depend on the following factors: (i) the specificities of the
interacting projectile and target, (ii) the particularities of the
secondaries emitted from a specific hadronic or nuclear interaction and (iii)
the magnitudes of the momentum transfers and of a phase factor
(with a maximum value of unity) in the rearrangement process in any collision. Collecting and combining
all these, we propose the relation to be given by \cite{pgr08}
\begin{equation}\displaystyle
N_R=4<N_{part}>^{1/3}\theta,
\end{equation}
where $<N_{part}>$ denotes the average number of participating
nucleons and $\theta$ values are the constants of proportionality and they are to be obtained phenomenologically
from the fits to the data-points.
\par
In order to study a nuclear interaction of the type
$A+B\rightarrow Q+ x$, where $A$ and $B$ are projectile and target
nucleus respectively, and $Q$ is the detected particle which, in the
present case, would be direct photon, the SCM
has been adapted to a fly over model. The fly-over model proposed here is on the basis of the suggestions from Wong \cite{wong}, the Glauber techniques and by using Wood-Saxon
distributions \cite{gorenstein}. The inclusive
cross-sections for direct photon production in different nuclear
interactions of the types $A+B\rightarrow {\gamma}_d+ x$ in the
light of this modified Sequential Chain Model (SCM) can then be written
in the following generalised form as \cite{pgr031}-\cite{pgr08}:
\begin{equation}\displaystyle
{E{\frac{d^3\sigma}{dp^3}}|_{AB\rightarrow{{\gamma}_d} x}=
a_{\gamma_d} {p_T}^{-N_R^{\gamma_d}} \exp(-\Delta^{\gamma_d}{p_T})}.
\end{equation}
 where $a_{\gamma_d}$, $N_R^{\gamma_d}$ and $\Delta^{\gamma_d}$ are
the factors to be calculated under certain physical constraints. With the details of
the calculations to be obtained from Refs.\cite{pgr031}-\cite{pgr08}, the
set of relations to be used for evaluating the parameters
$a_{\gamma_d}$ and $\Delta^{\gamma_d}$ is given below.
\begin{equation}\displaystyle{
a_{\gamma_d}=C_{\gamma_d}{\frac{3}{2\pi}}{\frac{(A \sigma_B + B
\sigma_A)}{\sigma_{AB}}}
{\frac{1}{1+a'(A^{1/3}+B^{1/3})}}{(\frac{1000}{s})}^{1/5}}
\end{equation}
and
\begin{equation}\displaystyle{
\Delta^{\gamma_d} \cong \xi s^{-0.3}}
\end{equation}
Here, in the above set of equations, the third factor of equation (5) gives a
measure of the number of wounded nucleons i.e. of the probable
number of participants, wherein $A\sigma_B$ gives the probability
cross-section of collision with `$B$' nucleus (target), had all
the nucleons of $A$ suffered collisions with $B$-target. And
$B\sigma_A$ has just the same physical meaning, with $A$ and $B$
replaced. Furthermore, $\sigma_A$ is the
nucleon(proton)-nucleus(A) interaction cross section, $\sigma_B$
is the inelastic nucleon(proton)-nucleus(B) reaction cross section
and $\sigma_{AB}$ is the inelastic $AB$ cross section for the
collision of nucleus $A$ and nucleus $B$. The values of
$\sigma_{AB}$, $\sigma_{A}$, $\sigma_{B}$ are worked here out in a
somewhat heuristic manner by the following formula \cite{na5002}
\begin{equation}
\displaystyle{ \sigma^{inel}_{AB} ~ = ~ \sigma_{0} ~
(A^{1/3}_{projectile} + A^{1/3}_{target} - \delta)^2}
\end{equation}
with $\sigma_{0} = 68.8$ mb, $\delta= 1.32$.
\par
Besides, in expression (5), the fourth term is a physical factor
related with energy degradation of the secondaries due to multiple
collision effects. The parameter $a'$ occurring in eqn.(5) above
is a measure of the fraction of the nucleons that suffer energy
loss. The maximum value of $a'$ is unity, while all the nucleons
suffer energy loss. This $a'$ parameter is usually to be chosen
\cite{wong}, depending on the centrality of the collisions and the
nature of the secondaries.
\par
The ``$a_{\gamma_d}$" factor in the expression (4) accommodates a wide range
of variation because of the existence of the large differences in
the in the normalizations of the direct photon cross-sections for
different types of interactions.
\par
It is clear from expression (6), that the parameter $\Delta$ has a
particular energy dependence given in a specific form with $\xi$
as a numerical factor arising out of the proportionality constant
and is treated here as a constant at a definite energy. For example, the value
of $\xi$ for direct photon production is taken here to be 10.33 at
$\sqrt{s_{NN}} = 200 GeV$.
By using Eqn.(4), we arrive at the expressions for production of
direct photons in $p+p$ and $d+Au$ (deuterium-gold)collisions at
energy $\sqrt s_{NN} =200 GeV$ at RHIC.
\begin{equation}
E\frac{d^3\sigma}{dp^3}|_{p+p\rightarrow\gamma_d +X}=
0.23\frac{1}{p_T^{2.56}}\exp(-0.430p_T)
\end{equation}
\begin{equation}
E\frac{d^3\sigma}{dp^3}|_{d+Au\rightarrow\gamma_d+ X}=
(8\times10^{-4})\frac{1}{p_T^{2.67}}\exp(-0.430p_T)
\end{equation}
The values of $N_R$ of eqn. (8) and (9) are calculated from
eqn.(3). The $<N_{part}>$ values for $p+p$ and $d+Au$ reactions are
taken to be 2 and $\sim179$.\cite{adler07}
\par
In figure (2), we have drawn
the solid lines depicting the model-based results with the help of
above two equations for $p+p$ and $d+Au$ interactions against the
experimental background \cite{adare1208}. Moreover, in that figure, the dotted line represents NLO pQCD-based results \cite{adare1208}. A point, in this context, needs to be addressed here. The disagreement observed in the low-$p_T$ by the SCM is due to the power law form of constituent rearrangement factor $1/p_T^{N_R}$. This dominance of this power law form disturbs the agreement between data and model-based calculations in ``soft" values. Whereas, for the case of NLO pQCD, the non-leading order-term in pQCD is responsible for better fitment of low-$p_T$ data and also of high-$p_T$ data.
\par
In a similar fashion, the invariant yields of direct photon in $Au+Au$ collision at RHIC energy $\sqrt s_{NN} =200 GeV$ for different centralities have been plotted in Fig.3. Data are taken from \cite{adare1205}. The solid lines in the figure show SCM-based results. We would like to mention one point in this regard. In our previous work, in Ref.\cite{pgr08}, we have used the RHIC-data published in the year 2004-2005. But, in this paper, we have used their recent data published in the year 2012. Obviously, there might be certain differences in the measurement and the reproduction of the experimental data reported by the PHENIX group.  \cite{adare1208}-\cite{adare1205}
\par
Sometimes, it is customary to write a relation between $E\frac{d^3N}{dp^3}$ and $\frac{1}{2 \pi p_T}\frac {dN}{dp_Tdy}$ in the following Fourier series form;\cite{voloshin}
\begin{equation}
E\frac{d^3N}{dp^3} ~ = ~ \frac{1}{2 \pi p_T}\frac {d^2N}{dp_Tdy}(1+\sum_{n=1}^{\infty}2v_n \cos(n(\phi-\Psi_{RP}))),
\end{equation}
where $\Psi_{RP}$ denotes the azimuthal angle of the reaction plane. The sine terms are not present because of symmetry with respect to the reaction plane. $v_1$ is referred to as directed flow and $v_2$ as elliptic flow. But in our model, we do not use the second term in the right hand side of eqn. (10) till date. For this reason, the collective phenomena in non-central nuclear collisions have not yet been studied by the SCM. Only the invariant yields of direct photon for different centralities in $Au+Au$ collisions at RHIC have been studied and the model-based results are given in Fig.(3).
\par
For $Pb+Pb$ collision at energy $\sqrt s_{NN} =2.76 TeV$ at LHC, the SCM-based expression for the invariant yield would be written as
 \begin{equation}
\frac{1}{2 \pi N_{evt}p_T}\frac {dN}{dp_Tdy}|_{Pb+Pb\rightarrow\gamma_d+ X}=
0.75\frac{1}{p_T^{3.33}}\exp(-0.505p_T)
\end{equation}
In Fig. 4, we have plotted the invariant yields for $Pb+Pb$ collision at energy $\sqrt s_{NN} =2.76 ~ TeV$ at LHC. Data are taken from Ref.\cite{wilde}. The solid line in the figure represents the SCM-based above equation (eqn.(11)) while the dotted and dashed lines show the pQCD and Thermal approaches respectively.\cite{sinha}
\par
There is yet another very important observable called nuclear
modification factor (NMF) $R_{AB}$ which for production of the
direct photon is defined \cite{isobe} by
\begin{equation}\displaystyle{
R_{AB}(p_T)=\frac{d^2N^{AB}/dp_Td\eta}{<T_{AB}(b)>d^2\sigma^{pp}/dp_Td\eta}}
\end{equation}
where the numerator is invariant direct photon yield in unit
rapidity in d+Au collisions and the denominator is the expected
yield from the $p+p$ collisions scaled by nuclear overlapping
functions $<T_{AB}(b)>=<N_{coll}(b)>/\sigma_{pp}$ for $d+Au$.
\par
the SCM-based results on NMF for  $d+Au$  and $Au+Au$ collisions for forward rapidities
are deduced on the basis of  Eqn.(12) and are given by the undernoted relations
\begin{equation}\displaystyle{
R_{AB}|_{d+Au\rightarrow \gamma+X}=1.20p_T^{-0.11},}
\end{equation}
\begin{equation}\displaystyle{
R_{AB}|_{Au+Au\rightarrow \gamma+X}=1.33p_T^{-0.15}.}
\end{equation}
In Fig. 5(a) and 5(b), we plot $R_{dAu}$  and $R_{AuAu}$ vs.
$p_T$ for $d+Au$ and  $Au+Au$ collisions respectively.
The solid lines in the figure show the SCM-based results against the
experimentally  measured results \cite{adare1208} and \cite{adare1205} respectively. The dotted line in the fig.5(a) represents the `Cronin+isospin' formulation \cite{adare1208} while in Fig. 5(b) it represents `Prompt+QGP' approach \cite{adare1205}.
\par
The very basic model used here is essentially of non-standard
type. Because, the detected `direct' photons are not produced here
by any sort of gluon-radiation as is believed in the framework and the
phrases of the standard model. On the contrary, we have identified
them to be the products from the decays of the vector-bosons,
viz., $\rho$ and $\omega$ mesons. This introduces the gross
non-standard element and a heretic viewpoint in the present work.
And this redefining has brought about a reversal of both physical
outlook and calculational approach. However, the theoretical
results obtained by the combination of the SCM, the fly-over model
and the Glauberian techniques are, on an overall basis, fairly in
agreement with the measured data for direct photon production at
RHIC and LHC. The results on nuclear modification
factor are also in accord with measurements.
\par
In the end, the main conclusions are: (i) The model for direct
photon studies presented here, which is a combination of power and exponential-laws, works modestly satisfactorily. (ii) The model proposed here for the production of direct photon comprises some well-measured coupling terms of $\rho-\omega-\pi$ and $\omega-\gamma$ and some other standard particle physics considerations. (iii) Selected comparisons of our model-based results reveal neither sharp disagreement with any of them, nor very good agreement with either of them which are generally of standard model variety.
\par
{\bf{ACKNOWLEDGEMENTS}}\\
The authors are thankful to the learned Referees for their constructive and helpful criticisms. They are also indebted to Prof. S. Bhattacharyya of Indian Statistical Institute for his kind help which, needless to mention, have benefited them a lot. The work is supported by University Grants Commission, India, against the order no.PSW-30/12(ERO) dt.05 Feb-13.

\newpage
\begin{figure}
\centering
\includegraphics[width=3.0in]{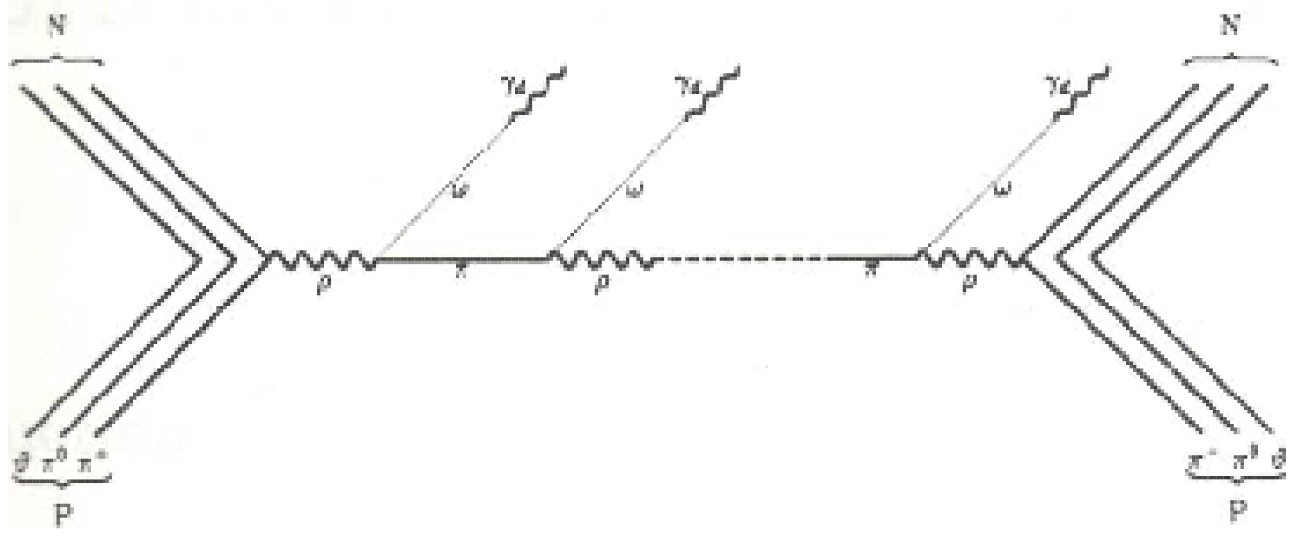}
\caption{\small Feynman diagram of direct photon production in Sequential Chain Model}
\end{figure}
\begin{figure}
\centering
\includegraphics[width=3.0in]{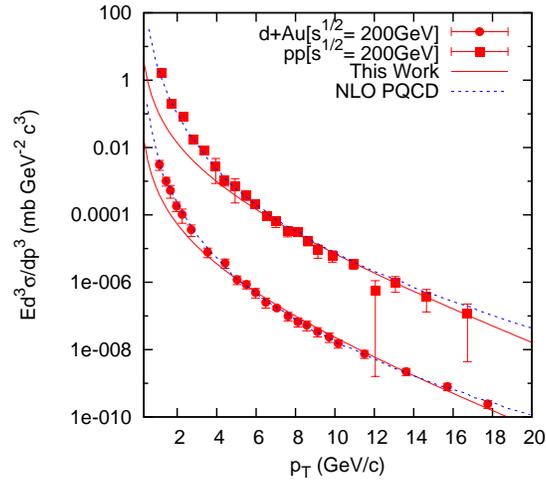}
\caption{\small Plot of the invariant
cross-sections for direct photon production in proton-proton and deuterium-gold
collisions at $\sqrt s_{NN}~ = ~200 GeV$ as function of $p_T$. The
data points are from \cite{adare1208}. The solid line
shows the SCM-based results while the dotted lie represents NLO PQCD-based output.}
\end{figure}
\begin{figure}
\centering
\includegraphics[width=2.8in]{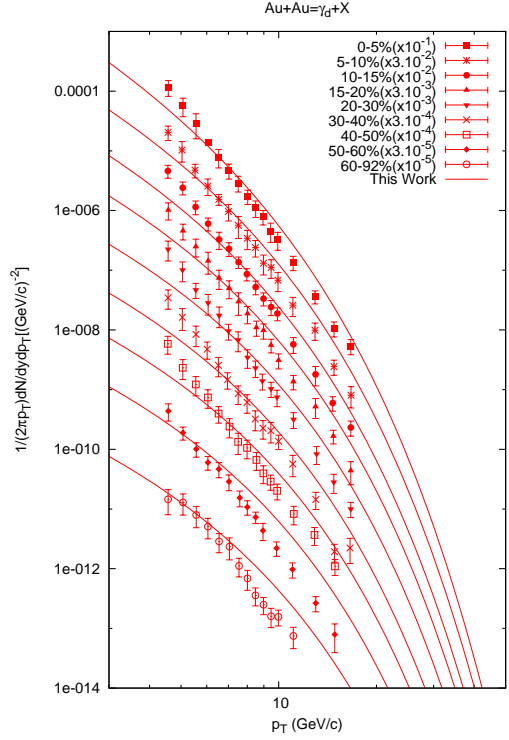}
\caption{\small Plot of the invariant
cross-sections for direct photon production for different centralities in gold-gold
collisions at $\sqrt s_{NN} ~= ~200 GeV$ as function of $p_T$. The
data points are from \cite{adare1205}. The solid lines
show the SCM-based results.}
\end{figure}
\begin{figure}
\centering
\includegraphics[width=2.8in]{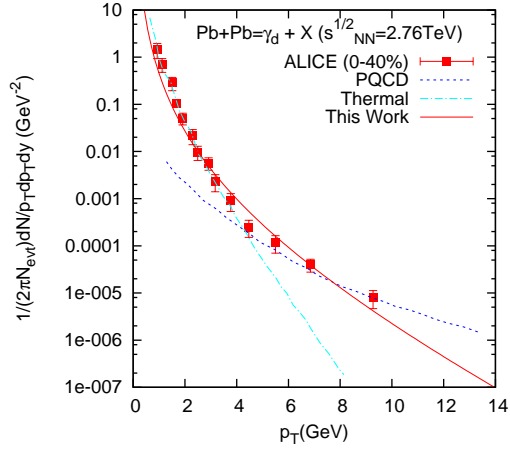}
\caption{\small Plot of the invariant
yields for direct photon production in lead-lead
collisions at $\sqrt s_{NN} =2.76 ~ TeV$ as function of $p_T$. The
data points are from \cite{wilde}. The solid line
shows the SCM-based results while the dotted and dashed lines show the PQCD and Thermal approaches respectively \cite{sinha}.}
\end{figure}
\begin{figure}
\subfigure[]{
\begin{minipage}{.5\textwidth}
\centering
\includegraphics[width=3.0in]{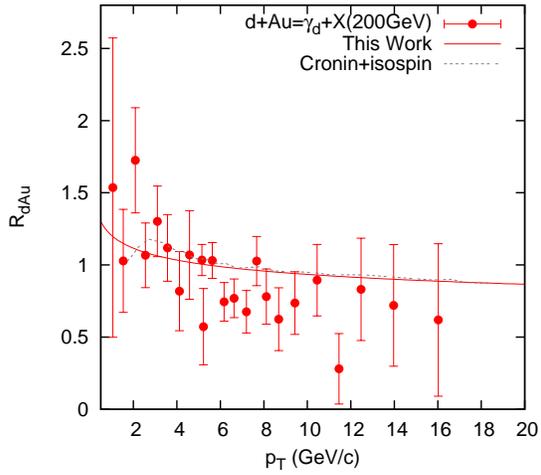}
\setcaptionwidth{2.6in}
\end{minipage}}%
\subfigure[]{
\begin{minipage}{0.5\textwidth}
\centering
\includegraphics[width=2.5in]{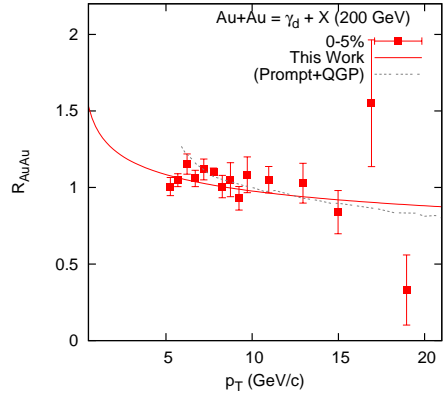}
  \end{minipage}}%
\caption{Plot of (a) $R_{dAu}$  and (b) $R_{AuAu}$(as defined in the text) vs. $p_T$-values.
Data are taken from \cite{adare1208} and \cite{adare1205} respectively. The solid lines in the figures show the SCM-based
theoretical results. The other lines in (a) and (b) are from the
Refs.\cite{adare1208}, \cite{adare1205} respectively.}
\end{figure}
\end{document}